\documentclass{article}
\usepackage{amsmath}
\usepackage{amsfonts}
\usepackage{amssymb}
\usepackage[version=4]{mhchem}
\usepackage{stmaryrd}
\usepackage{hyperref}
\hypersetup{colorlinks=true, linkcolor=blue, filecolor=magenta, urlcolor=cyan,}
\usepackage{bbold}
\usepackage{graphicx}
\usepackage[export]{adjustbox}
\usepackage{PRIMEarxiv}
\usepackage{comment}
\usepackage[utf8]{inputenc} 
\usepackage[T1]{fontenc}    
\usepackage{hyperref}       
\usepackage{url}            
\usepackage{booktabs}       
\usepackage{amsfonts}       
\usepackage{nicefrac}       
\usepackage{microtype}      
\usepackage{lipsum}
\usepackage{fancyhdr}       
\usepackage{graphicx}       
\graphicspath{{media/}}     
\usepackage{braket}
\usepackage{enumitem}
\newlist{steps}{enumerate}{1}
\setlist[steps, 1]{label = Step \arabic*:}
\title{Intermediate-qudit assisted Improved quantum algorithm for string matching with an Advanced Decomposition of Fredkin gate}

\author{
  Amit Saha$^{1*}$, Om Khanna$^{1, 2+}$ \\
  $^1$Atos, Pune, India\\
  $^2$Jai Hind College (Autonomous), University of Mumbai, Mumbai, India\\
  $^*$\texttt{abamitsaha@gmail.com} \\
  $^+$\texttt{aumkhanna@gmail.com}\\
}

\begin{document}
\maketitle

\begin{abstract}
  String-matching problem has a broad variety of applications due to its pattern-matching ability. The circuit-level implementation of a quantum string-matching algorithm, which matches a search string (pattern) of length $M$ inside a longer text of length $N$, has already been demonstrated in the literature to outperform its classical counterparts in terms of time complexity and space complexity.  Higher-dimensional quantum computing is becoming more and more common as a result of its powerful storage and processing capabilities. In this article, we have shown an improved quantum circuit implementation for the string-matching problem with the help of higher-dimensional intermediate temporary qudits.  It is also shown that with the help of intermediate qudits not only the complexity of depth can be reduced but also query complexity can be reduced for a quantum algorithm, for the first time to the best of our knowledge. Our algorithm has an improved query complexity of $O(\sqrt{N-M+1})$ with overall time complexity $O\left(\sqrt{N-M+1}\left((\log {(N-M+1)}  \log N)+\log (M)\right)\right)$ as compared to the state-of-the-art work which has a query complexity of $O(\sqrt{N})$ with overall time complexity $O\left(\sqrt{N}\left((\log N)^{2}+\log (M)\right)\right)$, while the ancilla count also reduces to $\frac{N}{2}$ from $\frac{N}{2}+M$. The cost of state-of-the-art quantum circuit for string-matching problem is colossal due to a huge number of Fredkin gates and multi-controlled Toffoli gates. We have exhibited an improved gate cost and depth over the circuit by applying a proposed Fredkin gate decomposition with intermediate qutrits (3-dimensional qudits or ternary systems) and already existing logarithmic-depth decomposition of $n$-qubit Toffoli or multi-controlled Toffoli gate (MCT) with intermediate ququarts (4-dimensional qudits or quaternary systems). We have also asserted that the quantum circuit cost is relevant instead of using higher dimensional qudits through error analysis.
\end{abstract}


\section{Introduction}

Quantum entanglement and superposition are two examples of quantum mechanical phenomena that are used in the idea of quantum computing for an asymptotic advantage \cite{chuang, preskil}. While the fundamental physics of quantum systems is not inherently binary, quantum computation is frequently stated as a two-level binary abstraction of qubits. However, higher dimensional systems can also be used to describe quantum processing. A qubit is expanded to a d-level or d-dimensional structure as a qudit \cite{Muthukrish-2000, Wang_2020}. In this article, an asymptotically improved binary circuit implementation of string-matching problem \cite{kmp}, has been addressed with temporary intermediate qutrits and ququarts by efficient decomposition of Fredkin gate \cite{barenco}. Since these only exist as intermediary states in a qudit system, where the input and output states are qubits, we can readily create a higher dimensional quantum state for temporary use by adding a distinct energy level \cite{Gokhale_2019}.

An essential family of algorithms known as "string-matching algorithms" looks for the location of one or more strings (also known as patterns) within a larger string or text.  These algorithms are used to discover answers for problems like text mining, pattern recognition, document matching, information security, network intrusion detection, and plagiarism detection.  When using exact matching, the pattern is precisely located within the text. The brute force algorithm is the most basic type of algorithm for finding a precise match in the string-matching problem. Let us see how it works: String $\mathcal{T}$ (to be searched) = ABCDEFGH and Pattern, $\mathcal{P}$ (to be matched) = CDEFG and P occurs once in $\mathcal{T}$: ABCDEFGH. With the brute force method, we merely attempt to match the first character of the pattern with the first character of the text. If we are successful, we move on to the second character and so forth. We move the pattern over one letter and attempt again if we run into a failure point. As a result, this method runs in $O(nm)$ time. However, the Knuth-Pratt-Morris algorithm, which has a worst-case temporal complexity of $\Theta(N+M)$, is the most well-known classical string-matching algorithm  \cite{kmp}. The most popular approximate string-matching algorithm also has a comparable run-time of $\Theta(N+M)$ \cite{approxstring}.

Quantum computing can be used to speed up string-matching algorithms. A precise string-matching quantum algorithm with $\tilde{O}(\sqrt{N}+\sqrt{M})$ query complexity was developed by Ramesh and Vinay \cite{ramesh}. In this method, each check is made using a nested Grover search to determine the location where a section of length $M$ from $\mathcal{T}$ matches the pattern  $\mathcal{P}$. However, this work does not create the specific oracles needed, and once we take into consideration the gate-level complexity of getting the text and pattern from a database, the total time complexity, expressed in units of gate depth, is bound to rise. For average-case matching, a different strategy for the dihedral hidden subgroup problem \cite{hidden} has a time complexity of $\tilde{O}\left((N / M)^{1 / 2} 2^{O(\sqrt{\log (M)})}\right)$ \cite{montanaro}.  The state-of-the-art work \cite{string} presents a string-matching algorithm, based on generalized Grover's amplitude amplification \cite{grover}, with a time complexity of $O\left(\sqrt{N}\left((\log N)^{2}+\log (M)\right)\right)$ along with $\frac{N}{2}+M$ ancilla for arbitrary text length $N$ and pattern length $M \leq N$. In this particular paper, we are also using the Grover-based string-matching algorithm to solve the string-matching problem, which achieves time complexity of $O\left(\sqrt{N-M+1}\left((\log {(N-M+1)}  \log N)+\log (M)\right)\right)$ with $\frac{N}{2}$ ancilla. We are using a system of intermediate qudits to implement a circuit that provides an asymptotic advantage over the state-of-the-art algorithm.

The main contribution of the article is summarized below:

\begin{itemize}
   
    \item We exhibit a first of its kind approach to implement an improved algorithm for the string-matching problem using a novel proposed decomposition of Fredkin gate using intermediate qutrit and multi-controlled Toffoli decomposition with intermediate ququart.
\item The proposed approach is sublimer with respect to the time complexity and space complexity with reduced ancilla qubits as compared to state-of-the-art approach \cite{string}.
    
    \item Our approach of solving string-matching problem outperforms the state-of-the-art approach \cite{string} with respect to circuit cost.
    
\end{itemize}

This paper has the following format. The background research required to carry out this suggested work is covered in Section \ref{preliminaries}. The circuit construction for the string-matching algorithm is proposed in Section \ref{construction} by decomposing the Fredkin gate using intermediate qutrits. The efficacy of the suggested approach in comparison to the state-of-the-art is analyzed in section \ref{analysis}. Our findings are summarized in Section \ref{conclusion}.

\section{Preliminaries}\label{preliminaries}

\subsection{A State-of-the-art Quantum Algorithm for String-matching}
The primary objective of string-matching algorithms is to find the location of a specific text pattern (P) within a larger string (S). The practical importance of these algorithms is in a wide variety of applications, from something as simple as searching for a particular word in a word processor to mapping DNA. 

In string-matching, we are given a long string S of length N, and our goal is to search for a pattern P contained in the string of length M, such that $M \leq N$. 
In Pradeep and Yunseong’s state-of-the-art paper \cite{string}, they constructed a quantum string-matching algorithm with a time complexity of $O(\sqrt{N}((\log(N)^2+\log(M))))$. The steps involved in their algorithm are as follows:

\begin{enumerate}
  \item It is based on the generalized Grover’s amplitude amplification technique. It works by initializing 2 quantum registers to store the bits of the target string of length $N$ and the pattern of length $M$. This process is done by using the identity and bit flip gates on 2 quantum registers ($\left|t_{0} t_{1} t_{2} \ldots t_{N-1}\right\rangle\left|p_{0} p_{1} \ldots p_{M-1}\right\rangle$, where $t_{i}$ and $p_{i}$ denote the ith bit of string $\mathcal{T}$ and pattern $\mathcal{P}$, respectively).
\end{enumerate}

\begin{enumerate}
  \setcounter{enumi}{1}
  \item The first register that contains the string $\mathcal{T}$ is changed into a combination of $N$ states, each of which is a bit-shifted version of the first register's initial state that has been moved by 0, 1, 2,..., $N-1$ bits. As a consequence, and presuming that the bit indices are stored in modulo-$N$ space,

\end{enumerate}

$$
\left(\frac{1}{\sqrt{N}} \sum_{k=0}^{N-1}\left|t_{0+k} t_{1+k} t_{2+k} \ldots t_{N-1+k}\right\rangle\right)\left|p_{0} p_{1} \ldots p_{M-1}\right\rangle
$$

\begin{figure}[ht]
    \centering
    \includegraphics[width= 10cm, height= 5cm]{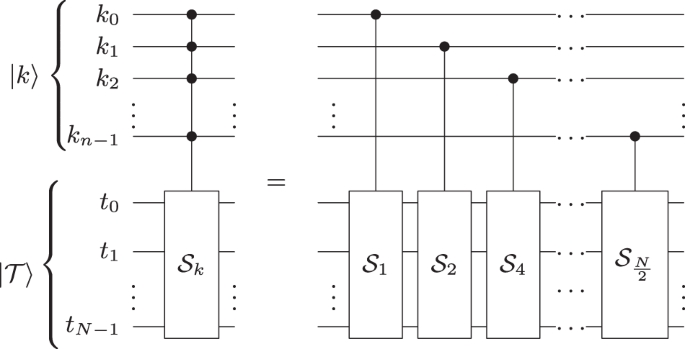}
    \caption{Circuit construction of cyclic shift operator \cite{string}.}
    \label{fig:CSO}
\end{figure}

This is done by a cyclic shift operator $S$ and the decomposition of the cyclic shift operator's circuit is shown in Figure \ref{fig:CSO}.

\begin{enumerate}
  \setcounter{enumi}{2}
  \item  Then, the XOR operation is performed on the first $M$ bits of the first register and the entire $M$ bits of the second register to obtain
\end{enumerate}

$$
\begin{gathered}
\frac{1}{\sqrt{N}} \sum_{k}\left|t_{0+k} t_{1+k} \ldots t_{N-1+k}\right\rangle \\
\left|\left(p_{0} \oplus t_{0+k}\right)\left(p_{1} \oplus t_{1+k}\right) \ldots\left(p_{M-1} \oplus t_{M-1+k}\right)\right\rangle .
\end{gathered}
$$

\begin{enumerate}
  \setcounter{enumi}{3}
  \item If the sequence matches the first $M$ bits of the $\mathcal{T}$, the second register contains only zeroes. If the text and the pattern vary by $d$ bit positions, the register holds $d$ ones.

  \item When looking for an exact match, the state where the second register contains only zeros or has fewer than $D$ matches using the generalized Grover search or amplitude amplification (in the case of fuzzy search), should be separated.

\end{enumerate}

A comparison on query and time-complexity between our work and other works \cite{ramesh, montanaro, string} is given in Table \ref{Tab1}. The oracles for \cite{ramesh, montanaro} offer arbitrary access to text and pattern bits. Because the execution time relies on the random-access oracles, which don't have a circuit-level design in the relevant papers, the time complexity for \cite{ramesh, montanaro} is unclear. In \cite{string} and our work, this random-access generator is not necessary. Instead, for the purposes of our work, an oracle is a Grover oracle that determines whether a register is in an all-zero state, similar to \cite{string}. Our detailed construction for such an oracle is discussed in this paper. We also follow the same algorithmic steps as \cite{string}. Albeit we design our circuit in such a way so that we achieve an asymptotic advantage over \cite{string} with the help of intermediate temporary qudits. Hence we directly compare our time-complexity with \cite{string}. The time complexity consists of three different parts, first is for the query, next is for the depth of the cyclic shift operator and the final part is for Grover's search. From Table \ref{Tab1}, it can be visualized that our proposed approach has an asymptotic advantage for the first two parts i.e.,  $\sqrt{N-M+1}$ and $\log {(N-M+1)}  \log N$ as compared to  $\sqrt{N}$ and $(\log N)^2$. For the last part, the complexity remains the same for the two approaches, but ancilla reduces to 0 from $M$.

\begin{table}[h!]
\centering
\resizebox{\textwidth}{!}{%
\begin{tabular}{llll}
\hline
Paper & Query complexity & Time complexity \\
\hline
\cite{ramesh} & $O\left(\sqrt{N} \log (\sqrt{N / M}) \log M+\sqrt{M}(\log M)^{2}\right)$  & - \\
\cite{montanaro} & $O\left((\sqrt{N / M}) 2^{(3 / 2)} \sqrt{\left(2 \log _{2} 3\right) \log _{2} M}(\log M)^{3 / 2} \log N\right)$  & - \\
\cite{string} & $O(\sqrt{N})$  & $O\left(\sqrt{N}\left((\log N)^{2}+\log (M)\right)\right)$ \\
This work & $O(\sqrt{N-M+1})$  & $O\left(\sqrt{N-M+1}\left((\log {(N-M+1)}  \log N)+\log (M)\right)\right)$ \\
\hline
\end{tabular}}
\caption{Comparison of our work with prior algorithms discussed \cite{ramesh, montanaro, string}.}
\label{Tab1}
\end{table}

\subsection{Toffoli Decomposition via Intermediate Qutrits}
Natural access to an infinite range of discrete energy levels is available to quantum processors. Therefore, using three-level qutrits is just an option to add another distinct energy level, but at the expense of allowing more space for error. Qutrits can replace the workspace provided by non-data ancilla qubits in typical circuits, allowing us to function more effectively. Qutrits are a 3-level quantum system where we consider the computational basis states: $\ket{0}$, $\ket{1}$ and $\ket{2}$. They are manipulated in a similar manner to qubits, however, there are additional ternary CNOT gates which may be performed on qutrits during the Toffoli decomposition. The Toffoli gate is the central building block of several quantum algorithms.  Since the Toffoli involves 3-body interactions, it cannot be implemented naturally in a real quantum devices. Usually, the Toffoli gate can be constructed by decomposing it into single and two qubit gates. For example CNOT gates require 6 such gates plus 7 $\mathrm{T}$ gates \cite{amy} as shown in Fig. \ref{toffoli_qubitonly}. Let’s look at the decomposition of the Toffoli gate using intermediate qutrits, since in this paper we are using Toffoli decomposition with intermediate qutrit:

\begin{figure}[h!]
\centering
\includegraphics[width=12cm, scale=1]{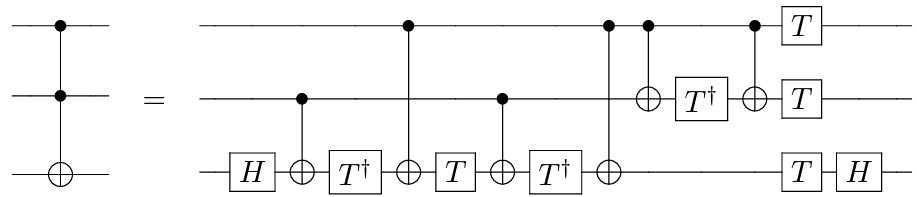}
\caption{Qubit-only Toffli decomposition with Clifford$+T$ gate-set \cite{amy}.}
\label{toffoli_qubitonly}
\end{figure}

\begin{figure}[h!]
    \centering
    \includegraphics[scale=.5]{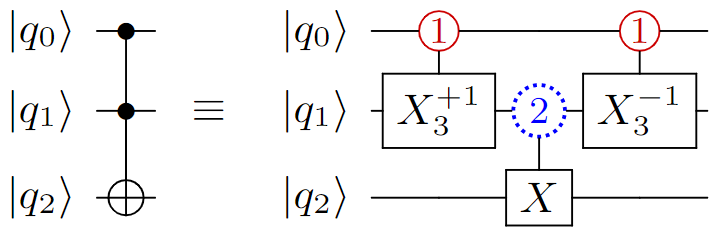}
    \caption{An example of Toffoli decomposition with intermediate qutrit, where input and output are qubits. The red controls activate on $\ket{1}$
and the blue controls activate on $\ket{2}$. The first gate temporarily elevates $q_1$ to $\ket{2}$ if both $q_0$ and $q_1$ were $\ket{1}$. $X$ operation is then only performed if $q_1$ is $\ket{2}$. The final gate acts as a miror of first gate and  restores $q_0$ and $q_1$ to their original state \cite{Gokhale_2019}}.
    \label{tof_qutrit}
\end{figure}

 In \cite{Gokhale_2019}, the authors demonstrated that we can momentarily inhabit the $\ket{2}$ state during the computation, making temporarily ternary. This circuit design can be integrated into any current qubit-only circuits because it maintains binary input and output.  Fig. \ref{tof_qutrit} depicts a Toffoli realization as seen through qutrits \cite{Gokhale_2019}. More precisely, the target qubit (third qubit) must undergo a NOT operation as long as the two control qubits are both $\ket{1}$. The first and second qubits are then subjected to a $\ket{1}$-controlled $X_{+1}$, where $+1$ stands for an increase of $1 \ (\text{mod}  3)$ to the target qubit. If and only if the first and second qubits were both $\ket{1}$, this raises the second qubit to $\ket{2}$. The target qubit is then subjected to a $X$ gate that is regulated by $\ket{2}$. As anticipated, $X$ is only performed when the first and second qubits were both $\ket{1}$. Lastly, a $\ket{1}$-controlled $X_{-1}$ gate cancels the impact of the first gate, returning the controls to their initial positions. The main result of this reduction is that the transient information can be stored in the $\ket{2}$ state from ternary quantum systems instead of ancilla. Therefore, three generalized ternary CNOT gates with a circuit depth of three are adequate to realize the Toffoli gate, in actuality, no $\mathrm{T}$ gate is needed. 

This Toffoli decomposition is further used to decompose the Fredkin gate for string-matching, which is thoroughly discussed in the next section. Before that, we have showcased the decomposition of multi-controlled Toffoli gate using ququarts, which is another important fundamental component for Grover's based string-matching. 

\subsection{Multi-controlled Toffoli Decomposition via Intermediate Ququarts}
In the previous section, we dealt with the construction of a Toffoli gate using a 3-level quantum system i.e., an intermediate qutrit. For the decomposition of $n$-qubit Toffoli gate,  the resources increase rapidly, requiring $O(n^2)$ two-qubit gates in qubit-only systems. However,
$n$-qubit Toffoli gates can be constructed efficiently using fewer resources than previous qubit-only designs with the help of intermediate qutrits \cite{Gokhale_2019}. 
Similar to this, there is research on the realization of $n$-qubit Toffoli with intermediary qudits; see \cite{Gokhale_2019, Nikolaeva2021, Nikolaeva2022, Nikolaeva2023, PhysRevA.105.062453}. Since this decomposition \cite{PhysRevA.105.062453} is more error-resistant and is the only one that can be scaled up to any finite dimensional quantum system as opposed to \cite{Gokhale_2019, Nikolaeva2021, Nikolaeva2022, Nikolaeva2023}, we are using it in this paper. It should be mentioned that the circuit cost of decomposition with intermediate qutrits \cite{Nikolaeva2022} and the decomposition with intermediate ququarts \cite{PhysRevA.105.062453} is comparable. Even so, not all quantum hardware supports the used gate-set from \cite{Nikolaeva2022}, and there is no error analysis for this decomposition because the error rates for the used gate-set and ternary systems are not documented in the literature. The gate-set utilised in \cite{Nikolaeva2022} is also not scalable to any finite dimensional system because it is not generalized to any such system.

\begin{figure*}[h!]
\centering
\includegraphics[scale=.7]{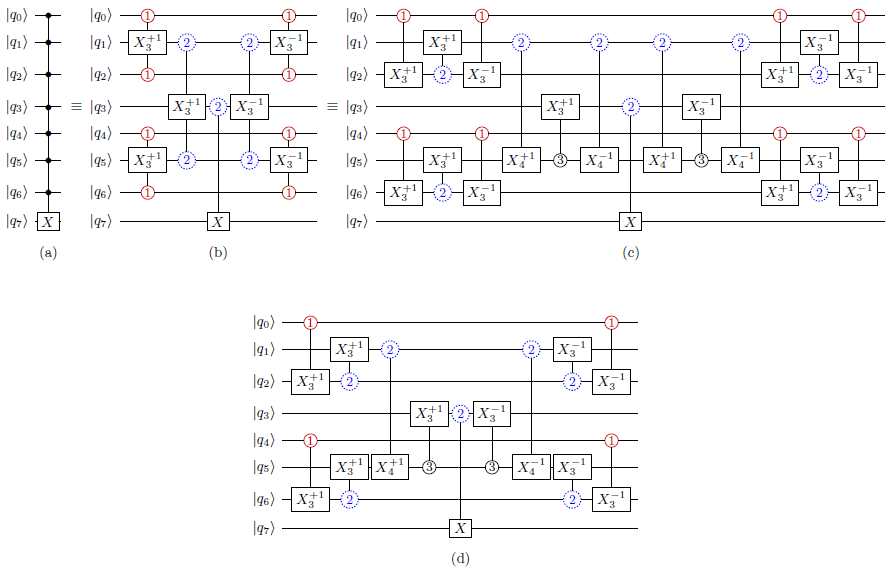}
\caption{(a) An 8-qubit Toffoli gate, (b) its decomposition in \cite{Gokhale_2019}, (c) its decomposition using a few ternary and/or quaternary CNOT gates in \cite{PhysRevA.105.062453}, and (d) its optimized decomposition in \cite{PhysRevA.105.062453}.}
\label{8qubit}
\end{figure*}

As an illustration, a multi-controlled Toffoli gate with 7 control qubits and 1 target qubit is taken into consideration, as shown in Fig. \ref{8qubit}(a). With the support of the Gokhale et al. \cite{Gokhale_2019} design, Fig. \ref{8qubit}(b) shows the realization of the generalized 8-qubit Toffoli gate as shown in Fig. \ref{8qubit}(a). In the same way that this method does, their circuit briefly saves information in the qutrit $\ket{2}$ state of the controls. However, they decompose their ternary Toffoli into 13 one-qutrit and two-qutrit gates, \cite{Di_2013} \cite{Gokhale_2019}, rather than saving temporary states in the quaternary $\ket{3}$ state. According to this method, three ternary and/or quaternary CNOT gates can be further reduced to two ternary and/or quaternary CNOT gates by using the identity rule, as shown in Fig. \ref{8qubit}(c) on Fig. \ref{8qubit}(d). The authors further decompose the ternary Toffoli into three ternary and/or quaternary CNOT gates using the $\ket{3}$. As a result, for a single Toffoli decomposition, this optimization can reduce the gate count from 13 to 2, and this method can also be applied to any dimensional quantum system.

This circuit design, as displayed in \ref{8qubit}(c) or \ref{8qubit}(d), can be understood as a binary tree of gates. More specifically, the circuit retains a tree structure with qubit inputs and outputs, and it has the characteristic that the intermediate qubit of each sub-tree and root can only be raised to $\ket{2}$ if all seven of its control leaves were $\ket{1}$. As a result, the circuit depth, where $n$ is the total number of controls, is exponential in $n$. Additionally, the overall number of gates is optimized because each quaternary qudit is acted on by a small constant number of two gates. The $n$-qubit Toffoli decomposition is novel because it uses a maximum of $2n - 3$ generalized CNOT gates ($n+1$ ternary CNOT gates and $n-4$ quaternary CNOT gates), which is less than the state-of-the-art. It is also novel because of its logarithmic depth optimization. This decomposition of the multi-controlled Toffoli gate has further played a vital role to reduce the query complexity and ancilla qubits of our proposed string-matching algorithm, which is discussed in the next section.

\section{Quantum Algorithm for String-matching with Intermediate Qudits}\label{construction}

\subsection{Our Proposed Fredkin Gate with Intermediate Qutrits}

In this section, we show an explicit circuit decomposition of the Fredkin gate using intermediate qutrits. Before that the state-of-the-art decomposition of a Fredkin gate with 7 CNOT and $7 \mathrm{~T}$ gates is discussed. Let's start with the circuit of Fredkin gate:

\begin{center}
\includegraphics[width=7cm, scale=0.7]{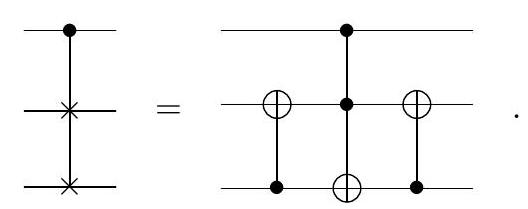}
\end{center}

Using the circuit identity, imported from Fig. \ref{toffoli_qubitonly} of \cite{string}, we find that the first CNOT gate in the Fredkin-gate circuit and the first two gates of the Toffoli-gate circuit forms a subcircuit. Thus, we obtain the state-of-the-art decomposition of Fredkin gate with Clifford $+\mathrm{T}$ gate set:

\begin{center}
\includegraphics[width=\textwidth, scale=1]{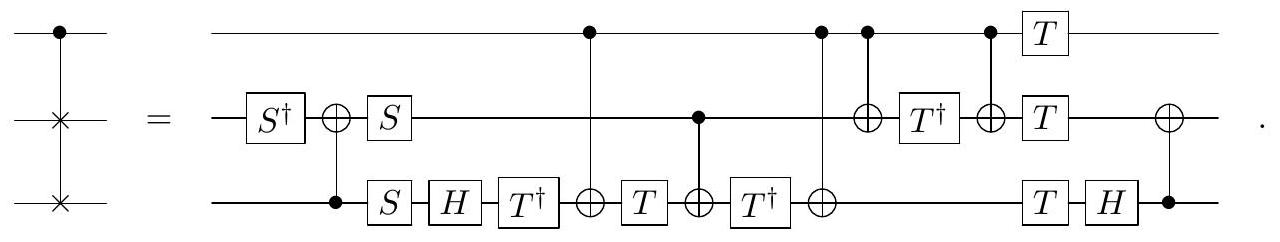}
\end{center}

\begin{figure}[h!]
    \centering  \includegraphics[width=8cm, scale=1]{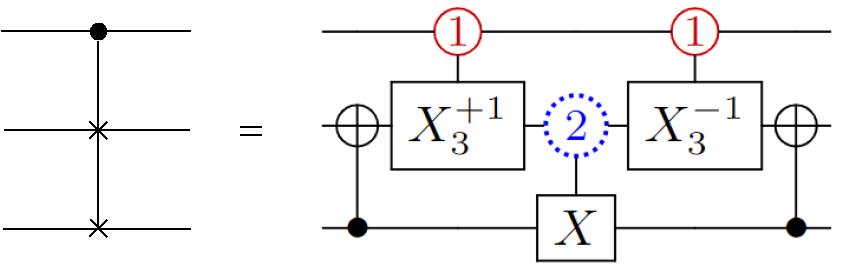}
    \caption{Advanced Fredkin gate with intermediate qutrit.}
    \label{adv_fredkin}
\end{figure}

The proposed Fredkin gate with intermediate qutrit is shown in Figure \ref{adv_fredkin}, where the Toffoli gate is decomposed as per Fig. \ref{tof_qutrit}. Only 2 CNOT gates and 3 ternary CNOT gates are required to construct this Fredkin gate, in fact no $\mathrm{~T}$ gate is required. We use this Fredkin gate further in our proposed string matching algorithm.

\subsection{Our Proposed Methodology for String-matching using Grover's Algorithm with Proposed Fredkin Gate}

\begin{figure}[h!]
    \centering
\includegraphics[width=14cm, scale=1]{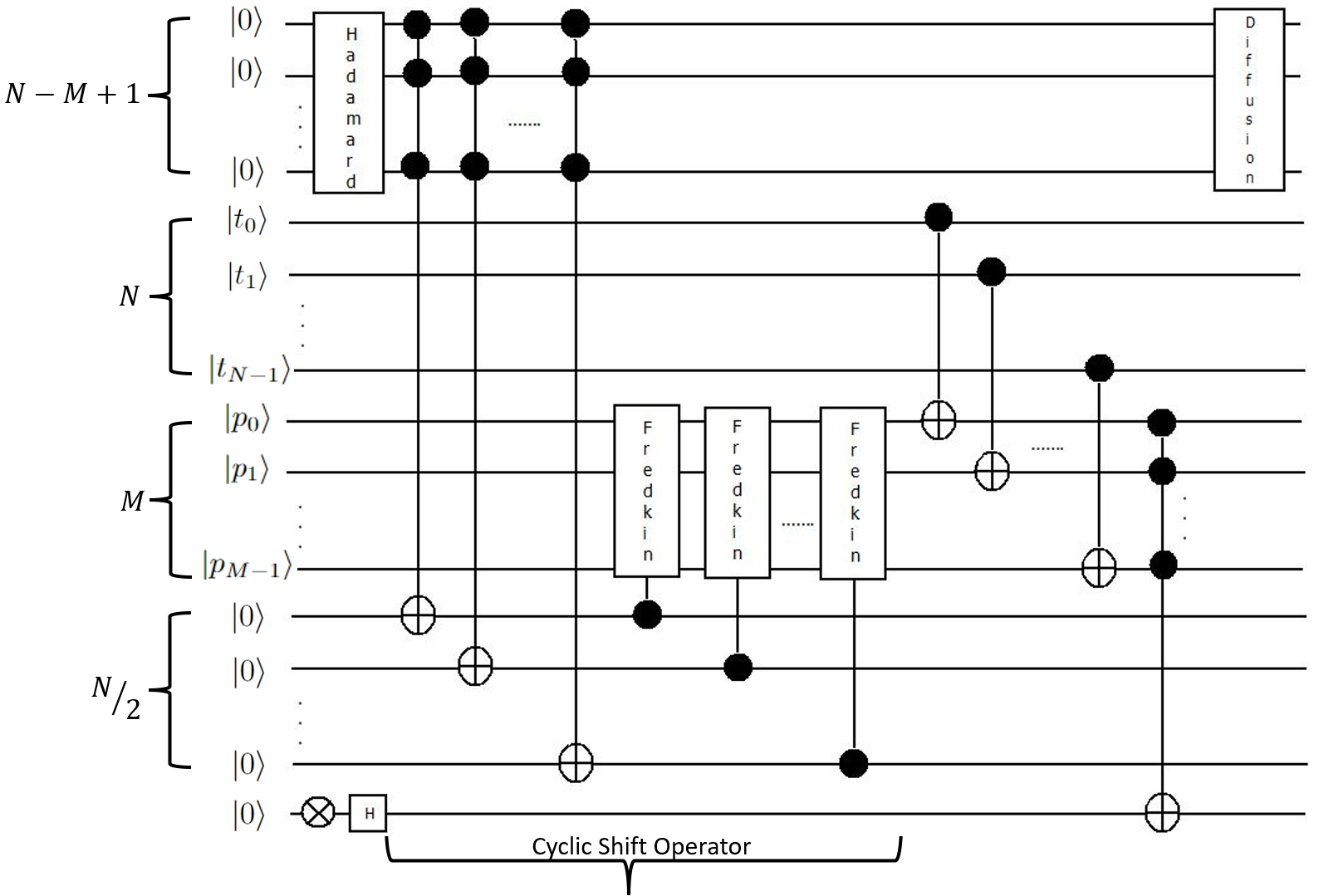}
    \caption{Complete circuit for proposed string-matching.}
    \label{completestring}
\end{figure}

We outline the proposed algorithm's thorough implementation in this section. We specifically describe the registers and transformations used to carry out the method.  The cyclic shift operator with proposed Fredkin gate is one of the key changes that will be used in our method compared to \cite{string}. Another key aspect of our algorithm is that due to the use of multi-controlled Toffoli decomposition with intermediate qudits, the query complexity has been reduced. We present the details of complete circuit construction for string-matching using Grover's algorithm, which is portrayed in Fig. \ref{completestring} for better visualization.

\begin{steps}

\item We also use quantum registers of $N$ and $M$ qubits, respectively, to encapsulate a binary string $\mathcal{T}$ of length $N$ and a binary pattern $\mathcal{P}$ of length $M$ as \cite{string}. To accomplish this, identity and bit-flip gates can be used on a quantum register with an initialization of $|0\rangle^{\otimes(N+M)}$. The encoded quantum state is as follows

$$
\begin{aligned}
|\mathcal{T}\rangle & =\left|t_{0} t_{1} \ldots t_{N-1}\right\rangle=\bigotimes_{i=0}^{N-1}\left|t_{i}\right\rangle \\
|\mathcal{P}\rangle & =\left|p_{0} p_{1} \ldots p_{M-1}\right\rangle=\bigotimes_{j=0}^{M-1}\left|p_{j}\right\rangle.
\end{aligned}
$$

 \item We now construct a composite initial state and an index register of $N-M+1$ qubits in the zero states, 

$$
|\psi\rangle=|0\rangle^{\otimes N-M+1}\left[\bigotimes_{i=0}^{N-1}\left|t_{i}\right\rangle\right]\left[\bigotimes_{j=0}^{M-1}\left|p_{j}\right\rangle\right]
$$

where, for ease of use, we assumed $N-M+1=2^{n}$. The index register is then subjected to a $n$ qubit Hadamard transform $H^{\otimes n}$ (Fourier transform in case of $N-M+1 \neq 2^{n}$ for $n \in \mathbb{N}$) to produce a uniform superposition of $|0\rangle,|1\rangle, \ldots|N-M\rangle$, 

$$
\left(H^{\otimes n}|0\rangle^{\otimes n}\right)\left[\bigotimes_{i=0}^{N-1}\left|t_{i}\right\rangle\right]\left[\bigotimes_{j=0}^{M-1}\left|p_{j}\right\rangle\right]=\left(\frac{1}{\sqrt{N-M+1}} \sum_{k=0}^{N-M}|k\rangle\right)\left[\bigotimes_{i=0}^{N-1}\left|t_{i}\right\rangle\right]\left[\bigotimes_{j=0}^{M-1}\left|p_{j}\right\rangle\right] .
$$

\item The next step is to use the cyclic shift operator $\mathcal{S}$, which left-circularly shifts the target state's qubits by $k$ places. $k$'s values are stored in the control state. The outcome of applying $\mathcal{S}$ to the first two registers is

$$
\begin{aligned}
& {\left[\mathcal{S}\left(\frac{1}{\sqrt{N-M+1}} \sum_{k=0}^{N-M}|k\rangle\right)\left(\bigotimes_{i=0}^{N-1}\left|t_{i}\right\rangle\right)\right]\left(\bigotimes_{j=0}^{M-1}\left|p_{j}\right\rangle\right)} \\
& =\frac{1}{\sqrt{N-M+1}} \sum_{k=0}^{N-M}|k\rangle\left(\bigotimes_{i=0}^{N-1}\left|t_{i+k}\right\rangle\right)\left(\bigotimes_{j=0}^{M-1}\left|p_{j}\right\rangle\right)
\end{aligned}
$$

Here, we provide a short explanation of the circuit design for the $\mathcal{S}$ cyclic-shift operator. We consider $k$ in its binary encoded form $\left|k_{0}\right\rangle\left|k_{1}\right\rangle \ldots\left|k_{N-M}\right\rangle$, such that $2^{0} k_{0}+2^{1} k_{1}+\ldots+2^{N-M} k_{N-M}=k$,  to implement the $k$-controlled circular shift operator $S_{k}$. Then, a combination of controlled-shift operators that shifts the target qubits by $k$ bits while depending on the $k$-controlled qubits can be used to execute the circular bitwise shift by $k$ in the second register. In other terms, a product of controlled shift operations can produce a shift of $k$ bits. We need the controlled-SWAP (Fredkin) gates to put the circular shift operator into practice. As an instance, a permutation of the form $P_{r}=\{N-r, N-r+1, N-r+2, \ldots, N-r-1\}$ is applied in modulo $N$ space by applying a cyclic shift operator $S_{r}$ by $r$ bits, where the $N-r$th bit is inserted in the zeroth position, the $N-r+1$th bit is inserted in the first position, and so on. Any one of these permutations can be realized into a series of transpositions. As a consequence, a cyclic shift operation can be realized into a SWAP operation's byproduct.

The number of SWAP-operation levels required to effectively implement the permutation is now determined. With a register having $N$ qubits, we can perform $\frac{N}{2}$ SWAP processes in parallel. We can transfer $\frac{N}{2}$ qubits to the appropriate locations in a single time step by using the $\frac{N}{2}$-parallel SWAP operator. Now we just need to arrange the remaining $N/2$ bits. The number of qubits that must be swapped drops by half at each succeeding time step. Therefore, using concurrent SWAP operations, we can arbitrarily permute $N$ qubits in $O(\log (N))$ time steps. This unitary process is illustrated diagrammatically with an example in Fig. \ref{swap}.  By using concurrent controlled-SWAP operators, shift operators can be implemented in $O(\log (\mathrm{N}))$ time steps.

\begin{figure}[h!]
    \centering
    \includegraphics[width=8cm, scale=1]{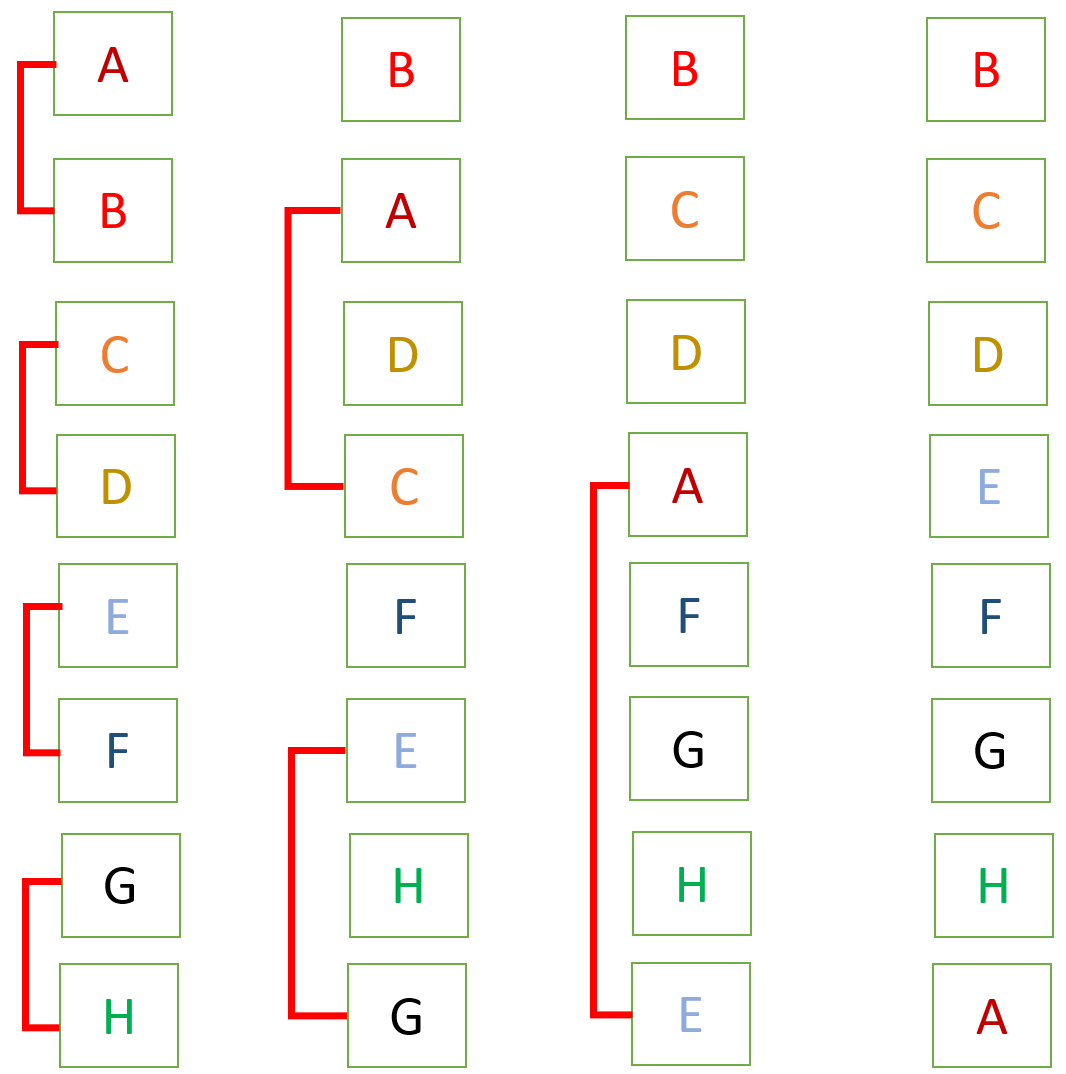}
    \caption{The cyclic-shift operator is shown in this diagram. In this case, we left-circularly shifted an 8-qubit register by one location over the course of three time steps. Generally speaking, this type of procedure can be carried out in depth $\log (N)$ using parallel SWAP operations, where $N$ is the size of the qubit register states.}
    \label{swap}
\end{figure}

Next, we go over how to use the same qubits in the index register to handle as many $\frac{N}{2}$ parallel swap processes. We succeed in doing this, at an expense of $\frac{N}{2}$ clean ancilla qubits. We start by considering a MCT operation, acting on the control qubits in a state $\left|k\right\rangle$ and $\frac{N}{2}$ clean ancilla qubits initialized to $|0\rangle$ as targets. This results in $\frac{N}{2}$ copies of $\left|1\right\rangle$, which can then be used to implement up to $\frac{N}{2}$ Fredkin gates in a single time step. Once all necessary Fredkin gates have been implemented, we undo the MCT operation and return all ancilla qubits to $|0\rangle$ for further operations. The time cost of the MCT operations with intermediate qudit is $O(\log (N-M+1))$ as index register of $\lceil\log(N-M+1)\rceil$ qubits are enough for our string matching since $N-M+1$ time cyclic shift operator is needed to be performed. For \cite{string}, they need index register of $N$ qubits as the logarithmic decomposition of MCT gates is not directly achievable in qubit-only circuits, hence they decompose their cyclic shift operator as shown in Fig. \ref{fig:CSO}. Since there are $O(\log (\mathrm{N}))$ parallel SWAP layers required for the implementation of the qubit permutation, the overall time complexity of cyclic shift operator is $O(\log (N-M+1)\log (\mathrm{N}))$.

\item At this juncture, we look to see if the pattern string kept in the third register matches the cyclically moved text strings in the second register. Each of the first $M$ bits in the second register and each of the $M$ bits in the third register are combined using an XOR operation. For instance, the sequences match if the XOR outputs are all zeros. Then, with the use of CNOT gates on a quantum computer, we acquire,

$$
\begin{aligned}
& \frac{1}{\sqrt{N-M+1}} \sum_{k=0}^{N-M}|k\rangle \text { CNOT }^{\otimes M}\left[\left(\bigotimes_{i=0}^{N-1}\left|t_{i+k}\right\rangle\right)\left(\bigotimes_{j=0}^{M-1}\left|p_{j}\right\rangle\right)\right] \\
& \quad=\frac{1}{\sqrt{N-M+1}} \sum_{k=0}^{N-M}\left[|k\rangle\left(\bigotimes_{i=0}^{N-1}\left|t_{i+k}\right\rangle\right)\left(\bigotimes_{j=0}^{M-1}\left|p_{j} \oplus t_{j+k}\right\rangle\right)\right] .
\end{aligned}
$$

For this purpose, the number of discrepancies between the pattern and the first $M$ bits of the string register is stored in the final register. In fact, if and only if those two string parts match exactly, it is all zero.

\item Finally, a Grover's oracle that works on the pattern register is necessary to finish our algorithm because it will amplify and help in the identification of exact matches or near matches. We can get this oracle in $O(\log (M))$ depth using novel decomposition of MCT gate using intermediate qudits without ancilla qubits. For a better understanding, we have given an example of proposed string-matching algorithm.

\end{steps}

\paragraph{Example:}

Let's take an example string  $\mathcal{T}$ (to be searched) = ABCDEFGH and Pattern, $\mathcal{P}$ (to be matched) = CDEFG. As per Fig. \ref{completestring}, $N=8$ and $M=5$. These $M$ and $N$ can be stated as $\ket{t}$ and $\ket{p}$ respectively. As per our proposed algorithm, we need an index register of $\lceil\log(N-M+1)\rceil$ i.e., $\lceil\log(8-5+1)\rceil=2$ extra qubits for cyclic shift operator as $\ket{k}$, which are initialized to 0. Next we need $\frac{N}{2}$ ancilla qubits as $\ket{a}$ for parallel Fredkin operation, which are also initialized as 0. Finally one output qubit for Grover's search with 1 as input. So the initial quantum state is:

$$\psi_0 \rightarrow \ket{k} \otimes \ket{t}  \otimes \ket{p} \otimes \ket{a} \otimes \ket{o}$$

$$\psi_0 \rightarrow \ket{00} \otimes \ket{ABCDEFGH} \otimes \ket{CDEFG} \otimes \ket{0000} \otimes \ket{1}$$

At first, we have to apply Hadamard transformation on first two qubits, hence the quantum state evolves as,

$\psi_1 \rightarrow \frac{1}{2}(\ket{00} \otimes \ket{ABCDEFGH} \otimes \ket{CDEFG} \otimes \ket{0000} \otimes \ket{1} +
\ket{01} \otimes \ket{ABCDEFGH} \otimes \ket{CDEFG} \otimes \ket{0000} \otimes \ket{1} + \ket{10} \otimes \ket{ABCDEFGH} \otimes \ket{CDEFG} \otimes \ket{0000} \otimes \ket{1} + \ket{11} \otimes \ket{ABCDEFGH} \otimes \ket{CDEFG} \otimes \ket{0000} \otimes \ket{1})$

Now, cyclic-shift operator comes into the action. When the value of index register $\ket{k}$ is $\ket{00}$, there will be no change in the systems. For the value of $\ket{01}$, there will be one place cyclic shift of $\ket{t}$. For that, through MCT operations ancilla register $\ket{a}$ becomes $\ket{1111}$ first,

$\psi_2 \rightarrow \frac{1}{2}(\ket{00} \otimes \ket{ABCDEFGH} \otimes \ket{CDEFG} \otimes \ket{0000} \otimes \ket{1} +
\ket{01} \otimes \ket{ABCDEFGH} \otimes \ket{CDEFG} \otimes \ket{1111} \otimes \ket{1} + \ket{10} \otimes \ket{ABCDEFGH} \otimes \ket{CDEFG} \otimes \ket{0000} \otimes \ket{1} + \ket{11} \otimes \ket{ABCDEFGH} \otimes \ket{CDEFG} \otimes \ket{0000} \otimes \ket{1})$

We now perform parallel Fredkin operations to shift one place of the string $\ket{t}$ for the index register $\ket{01}$ as shown in Fig. \ref{swap},

$\psi_3 \rightarrow \frac{1}{2}(\ket{00} \otimes \ket{ABCDEFGH} \otimes \ket{CDEFG} \otimes \ket{0000} \otimes \ket{1} +
\ket{01} \otimes \ket{BCDEFGHA} \otimes \ket{CDEFG} \otimes \ket{1111} \otimes \ket{1} + \ket{10} \otimes \ket{ABCDEFGH} \otimes \ket{CDEFG} \otimes \ket{0000} \otimes \ket{1} + \ket{11} \otimes \ket{ABCDEFGH} \otimes \ket{CDEFG} \otimes \ket{0000} \otimes \ket{1})$

We now again get back the value of ancilla qubits to $\ket{0000}$ through inverse operations so that further cyclic-shift operation can be performed for other indexed values,

$\psi_4 \rightarrow \frac{1}{2}(\ket{00} \otimes \ket{ABCDEFGH} \otimes \ket{CDEFG} \otimes \ket{0000} \otimes \ket{1} +
\ket{01} \otimes \ket{BCDEFGHA} \otimes \ket{CDEFG} \otimes \ket{0000} \otimes \ket{1} + \ket{10} \otimes \ket{ABCDEFGH} \otimes \ket{CDEFG} \otimes \ket{0000} \otimes \ket{1} + \ket{11} \otimes \ket{ABCDEFGH} \otimes \ket{CDEFG} \otimes \ket{0000} \otimes \ket{1})$

Similarly, we perform cyclic-shift operation for the other two index register's values, which are $\ket{10}$ and $\ket{11}$,

$\psi_5 \rightarrow \frac{1}{2}(\ket{00} \otimes \ket{ABCDEFGH} \otimes \ket{CDEFG} \otimes \ket{0000} \otimes \ket{1} +
\ket{01} \otimes \ket{BCDEFGHA} \otimes \ket{CDEFG} \otimes \ket{0000} \otimes \ket{1} + \ket{10} \otimes \ket{CDEFGHAB} \otimes \ket{CDEFG} \otimes \ket{0000} \otimes \ket{1} + \ket{11} \otimes \ket{DEFGHABC} \otimes \ket{CDEFG} \otimes \ket{0000} \otimes \ket{1})$

At this point, we perform the  XOR operation with the use of CNOT gates between first $M$ bits of $\ket{t}$ register and $M$ bits of $\ket{p}$ register. Outputs of $\ket{p}$ are all zeros for the indexed value of $\ket{10}$,

$\psi_6 \rightarrow \frac{1}{2}(\ket{00} \otimes \ket{ABCDEFGH} \otimes \ket{CDEFG} \otimes \ket{0000} \otimes \ket{1} +
\ket{01} \otimes \ket{BCDEFGHA} \otimes \ket{CDEFG} \otimes \ket{0000} \otimes \ket{1} + \ket{10} \otimes \ket{CDEFGHAB} \otimes \ket{00000} \otimes \ket{0000} \otimes \ket{1} + \ket{11} \otimes \ket{DEFGHABC} \otimes \ket{CDEFG} \otimes \ket{0000} \otimes \ket{1})$

We now perform the bit-flip operation through $X$ gate on $\ket{p}$ to get all ones,

$\psi_7 \rightarrow \frac{1}{2}(\ket{00} \otimes \ket{ABCDEFGH} \otimes \ket{CDEFG} \otimes \ket{0000} \otimes \ket{1} +
\ket{01} \otimes \ket{BCDEFGHA} \otimes \ket{CDEFG} \otimes \ket{0000} \otimes \ket{1} + \ket{10} \otimes \ket{CDEFGHAB} \otimes \ket{11111} \otimes \ket{0000} \otimes \ket{1} + \ket{11} \otimes \ket{DEFGHABC} \otimes \ket{CDEFG} \otimes \ket{0000} \otimes \ket{1})$

Next we perform Hadamard operation on output qubit $\ket{o}$ to perform the Grover's search,

$\psi_8 \rightarrow \frac{1}{2\sqrt{2}}[\ket{00} \otimes \ket{ABCDEFGH} \otimes \ket{CDEFG} \otimes \ket{0000} \otimes \ket{0} +
\ket{01} \otimes \ket{BCDEFGHA} \otimes \ket{CDEFG} \otimes \ket{0000} \otimes \ket{0} + \ket{10} \otimes \ket{CDEFGHAB} \otimes \ket{11111} \otimes \ket{0000} \otimes \ket{0} + \ket{11} \otimes \ket{DEFGHABC} \otimes \ket{CDEFG} \otimes \ket{0000} \otimes \ket{0} - (\ket{00} \otimes \ket{ABCDEFGH} \otimes \ket{CDEFG} \otimes \ket{0000} \otimes \ket{1} +
\ket{01} \otimes \ket{BCDEFGHA} \otimes \ket{CDEFG} \otimes \ket{0000} \otimes \ket{1} + \ket{10} \otimes \ket{CDEFGHAB} \otimes \ket{11111} \otimes \ket{0000} \otimes \ket{1} + \ket{11} \otimes \ket{DEFGHABC} \otimes \ket{CDEFG} \otimes \ket{0000} \otimes \ket{1})]$

We now perform the MCT operation between $\ket{p}$ and $\ket{o}$ and quantum state evolves as,

$\psi_{9} \rightarrow \frac{1}{2\sqrt{2}}[\ket{00} \otimes \ket{ABCDEFGH} \otimes \ket{CDEFG} \otimes \ket{0000} \otimes \ket{0} +
\ket{01} \otimes \ket{BCDEFGHA} \otimes \ket{CDEFG} \otimes \ket{0000} \otimes \ket{0} + \ket{10} \otimes \ket{CDEFGHAB} \otimes \ket{11111} \otimes \ket{0000} \otimes \ket{1} + \ket{11} \otimes \ket{DEFGHABC} \otimes \ket{CDEFG} \otimes \ket{0000} \otimes \ket{0} - (\ket{00} \otimes \ket{ABCDEFGH} \otimes \ket{CDEFG} \otimes \ket{0000} \otimes \ket{1} +
\ket{01} \otimes \ket{BCDEFGHA} \otimes \ket{CDEFG} \otimes \ket{0000} \otimes \ket{1} + \ket{10} \otimes \ket{CDEFGHAB} \otimes \ket{11111} \otimes \ket{0000} \otimes \ket{0} + \ket{11} \otimes \ket{DEFGHABC} \otimes \ket{CDEFG} \otimes \ket{0000} \otimes \ket{1})]$

We next perform the mirror operations to get back the quantum register $\ket{p}$ to its initial state,

$\psi_{10} \rightarrow \frac{1}{2\sqrt{2}}[\ket{00} \otimes \ket{ABCDEFGH} \otimes \ket{CDEFG} \otimes \ket{0000} \otimes \ket{0} +
\ket{01} \otimes \ket{BCDEFGHA} \otimes \ket{CDEFG} \otimes \ket{0000} \otimes \ket{0} + \ket{10} \otimes \ket{CDEFGHAB} \otimes \ket{CDEFG} \otimes \ket{0000} \otimes \ket{1} + \ket{11} \otimes \ket{DEFGHABC} \otimes \ket{CDEFG} \otimes \ket{0000} \otimes \ket{0} - (\ket{00} \otimes \ket{ABCDEFGH} \otimes \ket{CDEFG} \otimes \ket{0000} \otimes \ket{1} +
\ket{01} \otimes \ket{BCDEFGHA} \otimes \ket{CDEFG} \otimes \ket{0000} \otimes \ket{1} + \ket{10} \otimes \ket{CDEFGHAB} \otimes \ket{CDEFG} \otimes \ket{0000} \otimes \ket{0} + \ket{11} \otimes \ket{DEFGHABC} \otimes \ket{CDEFG} \otimes \ket{0000} \otimes \ket{1})]$

Finally we perform the Grover's amplitude amplification to obtain the final outcome,

$\psi_{11} \rightarrow -(\ket{10} \otimes \ket{ABCDEFGH} \otimes \ket{CDEFG} \otimes \ket{0000} \otimes \ket{1})$

The final quantum state depicts that the pattern '$CDEFG$' has a match in string '$ABCDEFG$'. The pattern can be found from the third position in the string. We can easily get the pattern's position by adding two with first position of the string since the indexed value suggests us as $10$ in binary i.e., two in integer. We also verify our results through simulation on the QuDiet platform \cite{qudiet}.


\section{Discussion}\label{analysis}

\subsection{Improved Time Complexity}
We determine our algorithm's time complexity in this subsection. Strings $\mathcal{T}$ and $\mathcal{P}$ require $O(1)$ time to encode. It also requires $O(1)$ time to apply the Hadamard or Fourier transform to the index register. The time required by the cyclic-shift operator  $\mathcal{S}$ is $O\left((\log {(N-M+1)} \log (N)\right)$. It takes time $O(1)$ to evaluate XOR outcomes using CNOT gates because they allow for simple concurrent processing. Last but not least, the complexity of the Grover oracle is $O(\log (\mathrm{M}))$. A single Grover step's complexity, which accounts for all the steps taken into account so far, is then $O\left(\log {(N-M+1)} \log (N)+\log (M)\right)$. The Grover steps must be repeated  $O(\sqrt{N-M+1})$ times in order for the Grover search to be successful. With this added intricacy, the total complexity is now $O\left(\sqrt{N-M+1}\left((\log {(N-M+1)} \log (N))+\log (M)\right)\right)$.

\subsection{Improved Space Complexity}
We also need $O(\log (N-M+1))$ qubits for the index register in addition to the $N$ and $M$ qubits required to store the search string and the pattern. We require $\frac{N}{2}$ ancilla qubits for the index register in order to implement our cyclic-shift operator in a depth-optimized manner.  We do not need any other extra ancilla qubit for our proposed approach. A comparative study of space complexity with \cite{string} is exhibited in Table \ref{Tab2}.

\begin{table}[h!]
\centering
\begin{tabular}{llll}
\hline
Space complexity & \cite{string} & This work \\
\hline
Data qubit & $N+M+\log N$  & $N+M+\log (\lceil{N-M+1}\rceil)$ \\
Ancilla qubit & $\frac{N}{2}+M$  & $\frac{N}{2}$ \\
\hline
\end{tabular}
\caption{Comparison of space complexity of our work with state-of-the-art algorithm \cite{string}.}
\label{Tab2}
\end{table}

\subsection{Improved Circuit Cost}
One can calculate the gate count in terms of CNOT and $\mathrm{T}$ gates according to the state-of-the-art circuit. Since it is widely anticipated that $\mathrm{T}$ gates will predominate the cost of implementation in the fault-tolerant regime, assuming the standard gate set of Clifford$+\mathrm{T}$, they chose those two gates as metrics. The best part of our proposed algorithm is no $\mathrm{T}$ gate is required for string-matching. We estimate the gate count in terms of CNOT, ternary CNOT and quaternary CNOT gates. It is important to remember the fact that the $n$-qubit Toffoli decomposition has a logarithmic depth and uses a maximum of $n+1$ ternary CNOT gates and $n-4$ quaternary CNOT gates. A comparative study of circuit cost with \cite{string} is exhibited in Table \ref{Tab3}.

\begin{table}[h!]
\centering
\resizebox{\textwidth}{!}{%
\begin{tabular}{llll}
\hline
Circuit cost & \cite{string} & This work \\
\hline
$\mathrm{T}$ & $(8 M-17+7(N-1) O(\log (N))) \times 2 \sqrt{N}$ & 0 \\
CNOT & $(7 M-12+(8 N-9) O(\log (N))) \times 2 \sqrt{N}$  & $((2N-1)O(\log (N))+M) \times 2 \sqrt{N-M+1}$ \\
ternary CNOT & 0 & $((N-M+2) O(\log (N-M+1))+(3N-1)O(\log (N))+(M-1)\log (M)) \times 2 \sqrt{N-M+1}$\\
quaternary CNOT & 0 & $((N-M-3) O(\log (N-M+1))+(M-4)\log (M)) \times 2 \sqrt{N-M+1} \times 2 \sqrt{N-M+1}$\\
\hline
\end{tabular}}
\caption{Comparison of circuit cost of our work with state-of-the-art algorithm \cite{string}.}
\label{Tab3}
\end{table}

The cost of the encoding step is zero because the strings $\mathcal{T}$ and $\mathcal{P}$ can be originally encoded in qubits in the $|0\rangle$ state using only the identity and $\operatorname{bit-flip}(X)$ gates. Hadamard gates are required for a Hadamard transformation of the index register which also requires zero cost. The stated permutation of size as large as $N$ can be divided into at most $N-1$ transpositions, so the cyclic shift operator $\mathcal{S}$ consists of an MCT gate with depth $O(\log (N-M+1))$ and at most $N-1$ Fredkin gates. As per our proposed Fredkin gate, each Fredkin gate costs 2 CNOT gates and 3 ternary CNOT gates. Thus the cyclic shift operator costs at most $(2N-1) O(\log (N))$ CNOT gates,  and  $((N-M+2) O(\log (N-M+1))+(3N-1)O(\log (N))$ ternary CNOT gates, and $((N-M-3) O(\log (N-M+1))$ quaternary CNOT gates. Next, the XOR operation requires $M$ CNOT gates. Lastly, the Grover oracle with multi-controlled Toffoli decomposition with intermediate qudits, can be implemented with $(M+1)\log (M)$ ternary CNOT gates and $(M-4)\log M$ quaternary CNOT gates without any ancilla. Lastly, for amplitude amplification, we need to repeat this $\sqrt{N-M+1}$ times. The total CNOT, ternary CNOT and quaternary CNOT count is, thus, given in Table \ref{Tab3} where the component of 2 comes from the necessity of applying a unitary to create the states $|\psi\rangle=$ $U|0\rangle$ and the inverse unitary $U^{\dagger}$ in order to amplify the amplitude.

\subsection{Error Analysis}

\begin{figure}[ht!]
    \centering
    \includegraphics[width=8cm, scale=1]{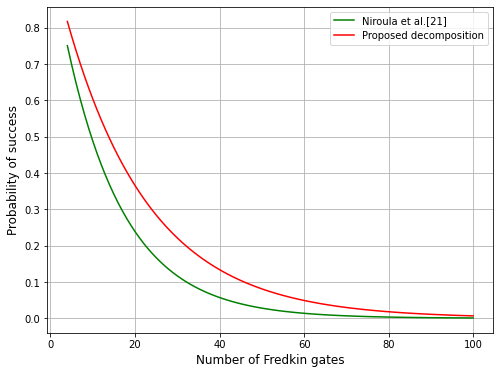}
    \caption{Probability of success for the decomposition of Fredkin gate using our proposed method (upper curve) versus the method in \cite{string} (lower curve).}
    \label{comp}
\end{figure}

Any quantum system is susceptible to different types of errors such as decoherence, noisy gates. For a binary quantum system, the gate error scales as $2^2$ and $2^4$ for 1- and 2-qubit gates respectively \cite{Gokhale_2019}. Furthermore, for qubits, the amplitude damping error decays the state $\ket{1}$ to $\ket{0}$ with probability $\lambda_1$. For a higher dimensional system, every state in level $\ket{i} \neq \ket{0}$ has a probability $\lambda_1$ of decaying. In other words, the usage of higher dimensional states penalizes the system with more errors. Nevertheless, the effect of these errors \cite{fisher} on the used decomposition of the Multi-controlled Toffoli gate has been studied by Saha et al. \cite{PhysRevA.105.062453} and the used decomposition of Toffoli gate for Fredkin gate has been studied by Gokhale et. al. \cite{Gokhale_2019}. They have demonstrated that even though the use of intermediate qudits results in a rise in error, the total error probability of the decomposition is lower than the ones used currently because there are fewer gates and less depth \cite{majumdar2021optimizing}. This interpretation is also applicable to our approach of solving the string-matching problem, since the gate cost and the depth have been reduced as compared to \cite{string}. Hence, we claim that our solution for the string-matching problems with intermediate qudits is superior in terms of error efficiency as compared to \cite{string}. The generalized Toffoli decomposition of \cite{PhysRevA.105.062453} that has been used in our proposed circuits for string-matching is also efficient with respect to crosstalk errors \cite{murli_crosstalk} due to its crosstalk-aware structure. Since the Fredkin gate decomposition is new to this paper, we show the probability of success for the Fredkin gate decomposition using the method of \cite{string} and our proposed method. As shown in Fig. \ref{comp}, we find that the decomposition in \cite{string} has a considerably higher error rate than the one we propose. This is explained by our decomposition's shallower depth and fewer gates. The advantage of our decomposition lies in the general substantial reduction in the gate count and the depth, despite the fact that some ternary and quaternary gates are used, which have a greater error probability due to the plague of dimensionality. Thus we can conclude that our circuit for string-matching is relevant instead of using higher dimensional qudits through error analysis.

\section{Conclusion}\label{conclusion}

We have built a quantum string-matching algorithm in this work that allows for a circuit-depth complexity of $O\left(\sqrt{N-M+1}\left((\log {(N-M+1)}  \log (N))+\log (M)\right)\right)$. Additionally, we offer a detailed gate-level version of our method, allowing for a precise calculation of the required quantum resources. This circuit for string-matching can be designed for any dimensional quantum system since the used gates are generalized, which makes the proposed approach generalized in nature. The primary use cases of the matching algorithm, such as a quick text search in a big file or spotting patterns in an image, can now be carried out more effectively. The proposed decomposition of Fredkin gate can be used in other algorithms for their efficient implementation. In fact, the overall findings show great promise for future work on effectively implementing other algorithms in intermediate qudit-assisted quantum computing. Whether the proposed approach will perform with similar efficiency in fault-tolerant regime \cite{majumdar2022fault} can only be answered with the evolution of more scalable
qudit-supported quantum hardware. Thus it is kept as a future aspect of this work when error correction for qutrits and ququarts would be feasible.

\section*{Acknowledgments}
There is no conflict of interest.

\bibliographystyle{abbrv}  
\bibliography{references}

\end{document}